\documentclass[cmap,twocolumn,amsmath,amssymb,amsfonts,10pt]{revtex4}
\usepackage[english]{babel}
\pdfoutput=1
\usepackage{hyperref}
\hypersetup{pdfauthor={E.V. Emelianov},pdftitle={IR Spectrometer Project for the
BTA Telescope}}
\usepackage{array}
\usepackage{ifpdf}
\ifpdf
        \usepackage[pdftex]{graphicx}
        \usepackage{cmap}
\else
        \usepackage[dvips]{graphicx}
\fi

\begin{document}
\selectlanguage{english}
\def\degr{\ensuremath{^\circ}}
\def\SPIE{{Society of Photo-Optical Instrumentation Engineers (SPIE)
Conf. Ser}}
\def\pasj{{Publ. Astron.~Soc. Japan}}
\def\pasp{{Publ. Astron.~Soc. Pacific}}
\newenvironment{pict}%
	{\begin{figure}[h]\begin{center}\noindent}{\end{center}\end{figure}}
\newenvironment{widepict}%
	{\begin{figure*}\begin{center}\noindent}{\end{center}\end{figure*}}
\newenvironment{tbl}%
	{\begin{table}\begin{center}\noindent}{\end{center}\end{table}}
\newenvironment{widetbl}%
	{\begin{table*}\begin{center}\noindent}{\end{center}\end{table*}}
\graphicspath{{./pic/}}
\def\thesection{\arabic{section}}
\def\thesubsection{\thesection.\arabic{subsection}}
\def\thesubsubsection{\thesubsection.\arabic{subsubsection}}
\def\m{.\hspace{-.2em}^m}
\def\'{.\hspace{-.2em}'}
\def\"{.\hspace{-.2em}''}

\def\HR#1{\protect\rule[0.2em]{0.6em}{#1}}

\title{IR Spectrometer Project for the BTA Telescope}
\author{V.~L.~Afanasiev}
\author{E.~V.~Emelianov}
\author{V.~A.~Murzin}
\affiliation{Special Astrophysical Observatory, Russian Academy of Sciences,
Nizhnii Arkhyz, 369167 Russia}
\author{V.~F.~Vdovin}
\affiliation{Institute of Applied Physics, Russian Academy of Sciences, Nizhny
Novgorod, 603950 Russia}
\affiliation{Nizhny Novgorod State Technical University, Nizhny Novgorod,
603950 Russia}

\maketitle

\section{Introduction}
The aim of the project is to extend the wavelength range accessible for
observations on the 6-m BTA telescope toward near infrared ($0.8\div2.5\,\mu
$m). This wavelength interval is productive for the study of ``cool'' objects
($T < 2000\,$K), dust enshrouded sources, and objects with redshifts up to $z =
10$. The  near infrared range contains lines that are of interest for spectral
diagnostics: rotational-vibrational bands of molecules important in astrophysics
(such as CO, OH, SiO, CH, CN, NH, H$_2$O, HCN, CH$_4$, etc.),
rotational transitions of H$_2$, numerous transitions of
neutral and ionized atoms. The list of astrophysical
tasks addressable with the 6-m BTA telescope in
the near IR is rather extensive and may include such
studies as:
\begin{itemize}
\item spectroscopy of protoplanetary nebulae;
\item search for exoplanets and brown dwarfs;
\item spectroscopy and photometry of regions of violent
star formation in our Galaxy and in the nearby and
distant ($z < 10$) galaxies;
\item stellar kinematics and morphology of systems of
various stellar populations in spiral galaxies;
\item physical conditions in ionized gas, stellar kinematics
and morphology in active galactic nuclei
(AGN) and starburst galaxies;
\item ultraluminous infrared galaxies (ULIRG);
\item determination of photometric redshifts from the
spectral energy distribution (SED).
\end{itemize}

\subsection{Specificities of the Design of IR Facilities}
\subsubsection{Photometry}
The Earth's atmosphere affects appreciably the IR radiation that passes through
it. Atmospheric absorption and scattering of IR radiation is quite significant
and, moreover, has an irregular nature, determined by
the spectral features of the absorption of IR waves by
the principal atmospheric gaseous components, and
first and foremost by water. Therefore photometric
observations in the IR are performed mainly in atmospheric ``transparency
windows''. The transmission
maxima of the I, J, H, and K-band filters coincide
with those of the Earth's atmosphere.

The so-called cold pupil stop is usually employed
in IR photometers to reduce thermal blackbody noise
from the telescope structures: a cooled diaphragm is
placed in the exit pupil plane to block such noise
sources as the primary focus cabin and the mirror
holder. All the other optical parts of the photometer
may be warmer, because their contribution to the total
noise is less important.

Cooling the exit pupil is impossible without extra
optics, and therefore when the instrument is operated
in the photometric mode a focal reducer has to be 
incorporated into the optical scheme even if the aperture
ratio is unchanged. To match the telescope scale with
the pixel size, it is better to reduce the image scale at
the detector, i.e., increase the aperture ratio.

As an example of an IR photometer in actual operation, let us consider FourStar,
a photometer with
a large field of view developed for the 6.5-m Magellan
Baade telescope of the Las Campanas observatory~\cite{2008SPIE.7014E..95P}.
The instrument has a~$10\'9\times10\'9$ field of view. Its
detector consists of a $2\times2$ array of HAWAII-2RG.
The collimator of the instrument also serves as the
entrance window. The image of the exit pupil produced by the collimator is
located immediately in front
of the camera. The main sources of thermal noise are
shielded by a cooled mask. The optical parts of the
camera are cooled to~$200\,$K. The filters, field correction
lenses, and detector located in the converging beam
after the camera are cooled to~$77\,$K. The filters are
placed on two six-slot filter wheels. The cryostat of
the instrument has the form a cylindrical structure
with a diameter and length of~$0.9$ and~$3\,$m, respectively. About~$45\,$l of
liquid nitrogen are needed to
maintain the operating temperature for 24~hours.

In 2005 the Special Astrophysical Observatory
acquired a CIRSI camera with two hybrid HAWAII-1
HgCdTe detectors to perform near-infrared photometry at the 6-m BTA telescope.
These detectors operate
in the $0.8\div2.5\,\mu$m wavelength range~\cite{2000SPIE.4008.1317M}.

The dewar with the camera, computer, and controller are located on a
$0.9\times0.9\times0.75\,$m$^3$ frame~\cite{1998SPIE.3354..431B}.
When assembled, the entire structure has a mass
of about 200\,kg. It can therefore be mounted only
on rather large telescopes. One detector covers
a~$2\'9\times2\'9$  sky area when operating in the primary
focus of the 6-m BTA telescope without a focal
reducer.

Numerous attempts to get CIRSI into working
condition have failed: cryostat leaks could not be
eliminated even to a degree that would allow laboratory tests to be performed.
CIRSI electronics have
become aged and obsolete: they were developed in
1997~\cite{1997SPIE.2871.1152B} and cannot interoperate with modern data
acquisition and reduction systems.

In our opinion, the most rational way to revive the
available IR equipment is to develop new hardware,
where the only parts left from CIRSI would be the
detectors.

\subsubsection{Spectroscopy}
IR spectroscopy is a relatively new trend in observational astronomy, which has
a great research potential. Interstellar extinction decreases rapidly toward
longer wavelengths and therefore infrared observations may provide data on
objects that are hidden by
optically thick dust and gas layers when observed in
the visible range. Furthermore, galaxies at large~$z$ can
be studied only in the IR because the main spectral
line features are redshifted.

The IR spectrographs currently used on major
telescopes suffer from many impassable limitations.
Among them is the small width of the spectral
interval covered by a single exposure. In addition,
high-resolution IR spectrographs are very expensive and highly specialized
instruments. Relaxing
the diffraction-limited resolution requirement would
make it possible to extend the application domain of
the instrument and make it more universal.

IR spectrographs actually differ little from their
visual counterparts except that they require deeper
cooling of the parts. Many published technical solutions can be found for such
spectrographs, however,
they are all essentially confined to focal reducers with
a grating or prism placed in the exit pupil to serve as a
dispersive element. Such a spectrograph is evidently
easy to convert into a photometer.

Mixed-mode instruments can operate in two
modes: as a photometer and as a spectrometer. For
example, for the Nasmyth focus of the 10-m Subaru
telescope a cooled IR-range camera/spectrograph
CISCO was developed~\cite{2002PASJ...54..315M}. The entire instrument is
housed in a cylindrical cryostat, where the pressure
and temperature are maintained at about $10^{-5}\,$Pa and
$54\div59\,$K, respectively. The instrument (when operating in the spectrograph
mode) has a spectral resolution of about $200\div250$ in the $0.8\div2.5\,\mu$m
wavelength
interval. The operation mode of the instrument can be
changed by replacing the elements (the prism, set of
grisms, and filter kit) installed on the stepper motor
controlled wheel. The readout noise and dark current
are sufficiently low to allow the instrument to be used
for faint objects ($22\m4$ and~$19\m7$ photometric and
and forspectroscopic observations, respectively).

The 1.88-m telescope of the Japanese Okayama
observatory is equipped with the IR camera/spectrograph OASIS, used both for
photometry and
long-slit spectroscopy~\cite{2000PASJ...52..931O}. The instrument employs
$256\times256\,$px HgCdTe NICMOS-3 arrays. In the
case of photometric observations the resolution is
$0\"97$/pixel, and when operated in the long-slit spectroscopy mode the
instrument provides a spectral
resolution $R=150\div1000$ (depending on the chosen grating: 300 or 75
lines/mm). The optical parts
of the instrument are cooled to 110\,K, whereas the
detector temperature is maintained at 80\,K (cooling
the instrument from room to operating temperature
takes about two days).

\begin{widepict}
\includegraphics[height=\textwidth,angle=270]{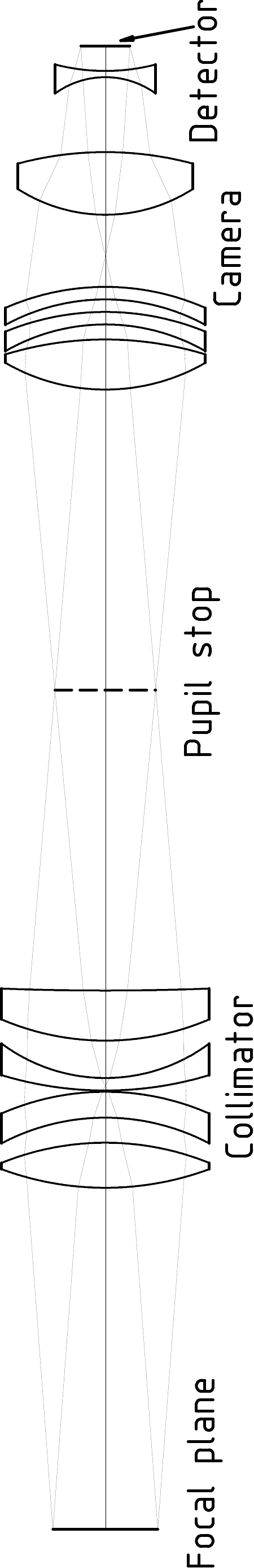}
\caption{Optical scheme of the IR focal reducer.}
\label{opt-sch}
\end{widepict}
\subsubsection{Detectors}
The operation of most of the near- and mid-infrared
detectors is based on the photoelectric effect. Charges
arising as a result of the internal photoelectric effect
can be accumulated in potential wells like in a CCD
cell, or the photo-conduction current created by these
charges can be recorded and amplified immediately.
Detectors can also be based on external photoelectric
effect.

Currently, two-dimensional IR detectors are mostly
based on HgCdTe chips manufactured using various
technologies. Let us list their main differences from
CCDs~\cite{2006sda..conf.....B}.
\begin{itemize}
\item Each pixel combines not only a cell to accumulate
the electrons emitted as a result of the internal
photoelectric effect, but also an amplifier. The by-effect is that the detector
gain varies from pixel to
pixel.
\item  In the readout process, the charge accumulated by
a CCD cell must pass through several other such
cells, whereas the signal from a single pixel of a
HgCdTe chip is read directly (via multiplexing of
outputs). Hence there is no charge transfer noise
in detectors of this type, resulting in reduced readout noise.
\item The CMOS multiplexer allows the signal to be
read off the cells of the HgCdTe array in arbitrary
order, which is beyond the capabilities of CCDs.
\item Another advantage of detectors based on CMOS
multiplexers is that they allow nondestructive signal readout: the current state
of the charge accumulated by array cells can be read out at any time
during exposure without fatal distortion.
\end{itemize}
Since the free charge carriers form in the semiconductor not only when induced
by photons, but
also due to thermal energy, the detector has to be
cooled to reduce noise. The operating temperature of
the detector decreases with increasing wavelength.
The operating temperature limit of the detector
can be approximately computed by the formula
$T_{max}=\frac{200}{\lambda_{max}}$,
where temperature is in Kelvin and the
long-wavelength cut-off is in microns.

Given that all the properties of the
CMOS multiplexer are sensitive to its operating
temperature, the optimum temperature should be
chosen carefully for each detector and maintained
with high stability throughout the entire exposure.

The $1024\times1024\,$px HAWAII-1 HgCdTe detector
array (with a pixel size of $18.5\times18.5\,\mu$m) acquired by
the Special Astrophysical Observatory has its max-
imum quantum efficiency (60\% at $T=78\,$K) in the
 $1.9\div2.5\,\mu$m wavelength interval. The depth of the po-
tential well of the detector cells is~$100\,$Ke$^-$. The gain
in the original readout system is about 8\,e$^-$/ADU.
The detector response curve is linear to within 1\% in
the $3000\div7000$\,ADU interval~\cite{CIRSI}.

HAWAII arrays are combined instruments with
an individual preamplifier for each element, resulting
in a highly increased pixel-by-pixel nonuniformity of
the detector. This creates additional challenges at the
stage of reduction of the acquired data. Special image
reduction techniques (correlated double sampling at
readout, post-readout reduction of the array of intermediate images) can be used
to increase the dynamic
range of the detector.

\begin{widepict}
\includegraphics[width=\columnwidth]{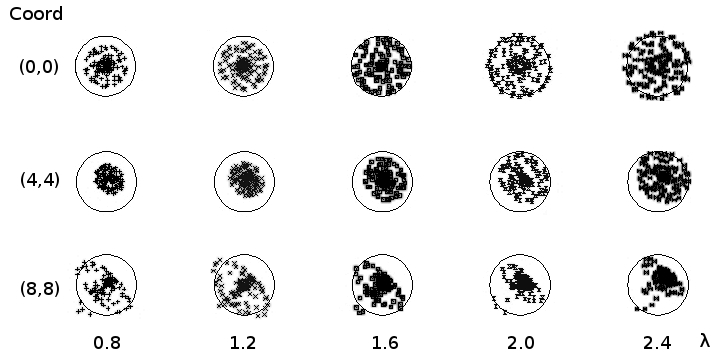}\hfill
\includegraphics[width=\columnwidth]{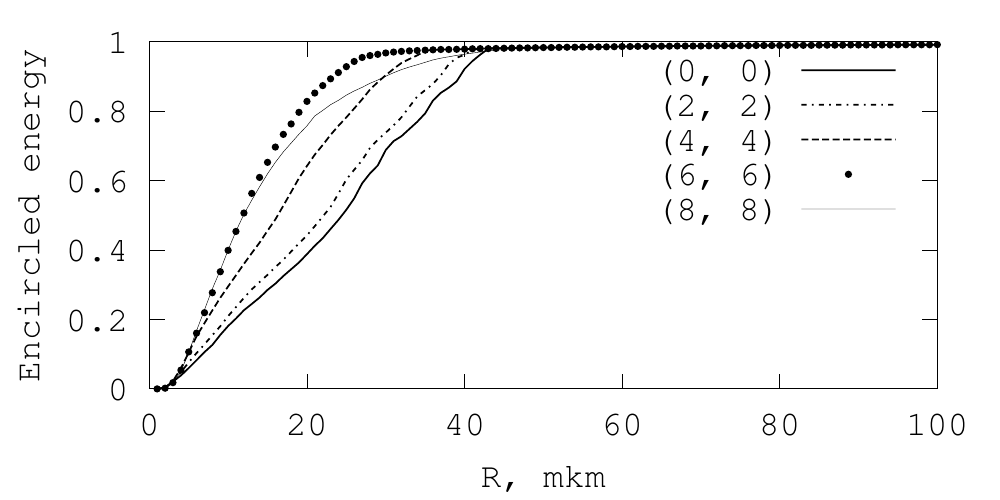}
\caption{The computed spot diagrams of the focal reducer (left). The size of the
circle on the diagram is $1''$ ($75\,\mu$m). The field
coordinates for the circle of confusion are given in the detector space
(in~mm). Wavelength is in~$\mu$m. The right-hand panel
shows the encircled energy (field coordinates are in~mm).}
\label{opt_characteri}
\end{widepict}
\subsubsection{Optical Components}
Given that most of the materials used in visual optics
are to a certain degree opaque for infrared radiation,
new IR materials have to be studied and implemented.
High-resistivity silicon is the classical material used
in the infrared: it is a replacement of sorts for glass
that is employed in the visible range. Also used in the
IR is chalcogenide glass (various metal compounds
with elements of the sixth periodic group--oxygen,
sulphur, selenium, tellurium, and polonium). Most of
these compounds are transparent for radiation with
wavelengths up to $11\div12\,\mu$m. Composite glasses
allow this wavelength range to be further expanded to
$18\div20\,\mu$m.

IR filters are often made using the same techniques as those employed for the
visible range. For
example, absorption cells or multiple reflection filters
may be used as broadband filters. Fabry-Perot etalons
can be used to filter narrow IR bands. Broader photometric bands can be created
with interference filters.

The rather large width of the near IR spectral
range ($0.8\div2.5\,\mu$m) combined with the resolution of
$R=1000\div3000$ prevents making an efficient spectrograph with a single
diffraction grating. However,
the use of a set of changeable gratings complicates
the design and increases the size of the cryostat, ultimately raising the cost
of the entire instrument. Furthermore, classical ``ruled'' gratings are a bad
choice
for the infrared, because (1) their efficiency decreases
sharply with decreasing grating period (thereby reducing the limiting resolution
of the instrument) and
(2) the grating itself becomes very unwieldy (otherwise the diffraction
efficiency of the instrument would
decrease even further because of the too wide instrumental function).

Volume Phase Holographic Gratings (VPHG)~\cite{2000PASP..112..809B}
have been used extensively in astronomical observations in recent years. Unlike
classical gratings, which
use spatial modulation, VPHG also uses modulation of the refraction index within
a thin layer of
dichromated gelatin of which the VPHG is made.
The gelatin layer must be thicker than the maximum
wavelength of the considered spectral interval.

The physics of light diffraction on VPHG is similar
to the mechanism of X-ray scattering on crystal lattices. VPHG reach maximum
efficiency if the Bragg
conditions are satisfied, which is equivalent to mirror
reflection from planes of constant refraction index.
The base wavelength satisfying the Bragg condition
can be changed by varying the angle of light incidence
onto the grating. Hence the maximum diffraction
efficiency can be achieved over a sufficiently wide
wavelength interval by changing the Bragg angle.
The maximum quantum efficiency of a VPHG may be
as high as $90\div95\%$ even if the radiation wavelength
is comparable to the grating period (the efficiency of
common gratings is known to decrease sharply in
this case). Other advantages of VPHG include weak
scattering and weak dependence of the grating efficiency on the polarization of
the radiation flux studied.

\section{Optical Arrangement of the Spectrometer}
The proposed optical arrangement (Fig.~\ref{opt-sch}) includes a four-lens~$F/4$
collimator with a focal length
of~$F=160\,$mm, and a five-lens~$F/2$ camera with a
focal length of~$F=100\,$mm and a  $2\omega=16\degr$ field of
view. The collimator corrects the coma and astigmatism of the parabolic mirror
of the 6-m BTA telescope
($F=24000\,$mm, $F/4$). The exit pupil relief and diameter are equal to 120 and
40\,mm, respectively, and the
back focal distance of the camera is equal to 10\,mm.

The optical arrangement of the instrument allows
it to operate in two modes: photometric and spectroscopic. The optical parts
are made of BaF$_2$ , CaF$_2$,
ZnSe, and SiO$_2$~(fused quartz, which
is transparent in the IR) optical crystals. All surfaces are covered with
anti-reflective coating optimized for the $0.8\div2.5\,\mu$m
wavelength interval.

Some of the chalcogenides used to make the
optical parts of the instrument are rather fragile
and therefore the corresponding lenses are mounted
on individual protective holders, which, in turn, are
mounted on the common holder of the collimator and
camera.

According to our computations, the secondary
spectrum of the focal reducer does not exceed
$\pm150\,\mu$m in the $0.8\div2.5\,\mu$m wavelength interval, and
the chromatic aberration of position does not exceed
1\,px (about $20\,\mu$m) at the edge of the field of view
(see Fig.~\ref{opt_characteri}). Aberrations in the field do not exceed the
size of two detector pixels (about $40\,\mu$m). The circle
of confusion over the detector field ($\pm9.47\,$mm, $\pm2\'3$)
for the operating wavelength interval fits within one
arcsecond (about $75\,\mu$m).

To reduce the length of the instrument, diagonal
mirrors are mounted in the parallel beam of the focal
reducer. The main feature of the instrument that distinguishes it from its
analogs is that it allows choosing the operating wavelength interval by
adjusting
the angle of incidence of the light beam onto the
diffraction grating. To this end, the tilt angle of the
diagonal mirrors is changed synchronously with the
shift of the grating (via a sine drive mechanism).

The parameters of the lenses of the focal reducer
are listed in Table~\ref{lens}. The optics are designed to op-
erate at a temperature of~$-40\degr$C. Simulations of
the behavior of the focal reducer in the case of small
deviations of the temperature from its design level
showed the image properties to be sufficiently stable
in the~$-40\degr\pm5\degr$C interval: only the focus of the entire
system changes (and this change can be compensated by shifting the detector).

\begin{widetbl}
\def\arraystretch{1.2}
\begin{tabular}{|>{\centering}m{7mm}|*{6}{>{$}c<{$}|}c|}
\hline
\bf No. &\mathbf{R_1},\text{ mm} &\mathbf{R_2},\text{ mm} &
\mathbf{H},\text{ mm} &\mathbf{X},\text{ mm} &\mathbf{D},\text{ mm} &
\mathbf{K},\,\mathbf{10^{-6}\mathrm{K}^{-1}} & \bf Material\slash
comments\\\hline
FP  & - & - & - & 136.46 & 32 & - & \small focal plane\\
1 & 119.88 & -110.84 & 18.00 & 10.00 & 83 & 18.9 & CaF$_2$\\
2 & -89.22 & -103.61 & 10.34 & 0.50 & 83 & 7.1 & ZnSe\\
3 & 143.21 & 69.60 & 5.00 & 14.93 & 83 & 0.5 & SiO$_2$\\
4 & 106.10 & 1104.77 & 20.00 & 120.00 & 83 & 18.4 & BaF$_2$\\
PS& - & - & - & 120.00 & 40 & - & \small exit pupil \\
5 & 80.23 & -137.16 & 20.00 & 6.00 & 80 & 18.4 & BaF$_2$\\
6 & -80.37 & -105.66 & 5.00 & 5.00 & 80 & 0.5 & SiO$_2$\\
7 & -83.63 & -99.53 & 5.00 & 28.83 & 80 & 7.1 & ZnSe\\
8 & 64.73 & -135.61 & 25.00 & 29.92 & 70 & 18.4 & BaF$_2$\\
9 & -34.30 & 80.14 & 2.50 & 10.00 & 40 & 0.5 & SiO$_2$\\
D & - & - & - & - & 19 & - &\small detector\\
\hline
\end{tabular}
\caption{Parameters of the focal reducer. No.~-- designation of a surface or the
number of a lens; $R_1$ and $R_2$~-- the curvature
radii of the lens surfaces; $H$~-- the lens thickness; $X$~-- the distance
between the neighboring surfaces; $D$~-- the lens
aperture, and $K$~-- the thermal expansion coefficient. All quantities are
given for the temperature of~$-40\degr$C.}
\label{lens}
\end{widetbl}

\subsection{Photometric Mode}
\begin{figure}
\hbox to\columnwidth{\hfill\includegraphics[width=.35\columnwidth]{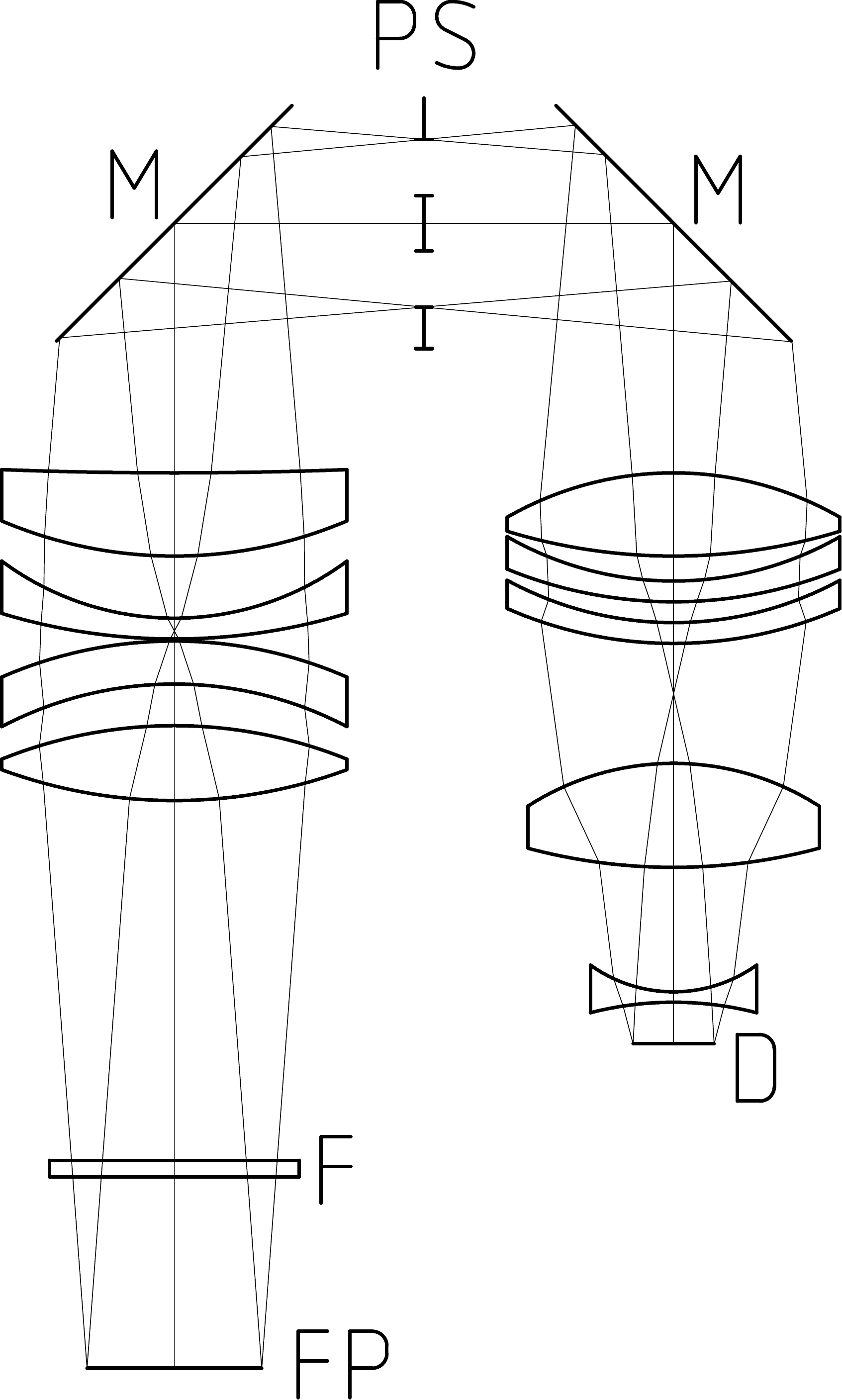}
\hfill\includegraphics[width=.35\columnwidth]{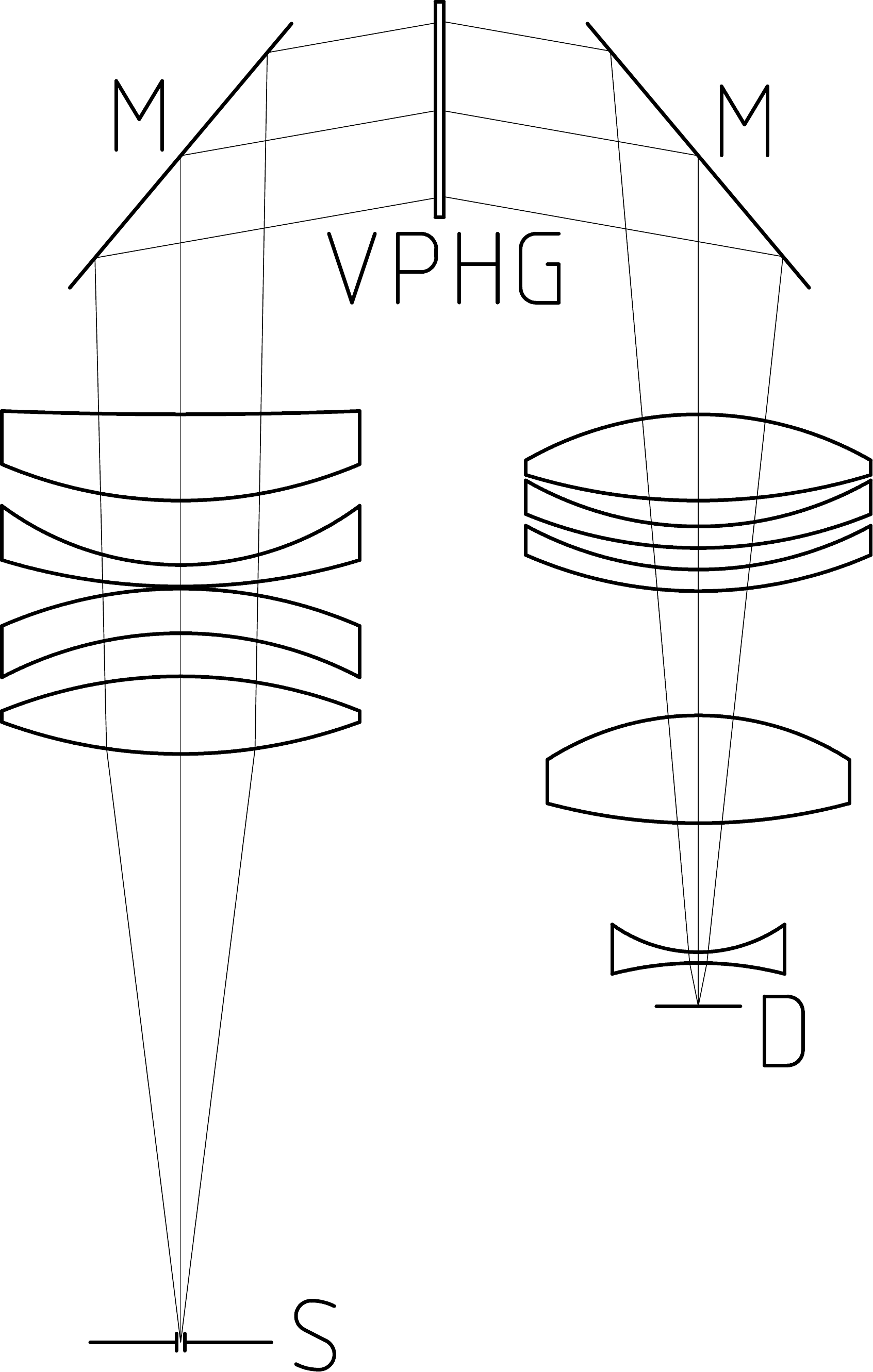}\hfill}
\caption{Diagram of the IR reducer operating in the pho-
tometric (left) and spectroscopic (right) modes. Here M
are the diagonal mirrors; FP is the focal plane of the
telescope; F is the filter; PS is the exit pupil stop; D is
the detector; S is the slit, and VPHG is the diffraction
grating.}
\label{modes}
\end{figure}

\begin{widetbl}
\begin{tabular}{|p{0.55\textwidth}|c|c|c|c|}
\hline
Filter & I & J & H & K\\
\hline
Central filter wavelength $\lambda_0$, $\mu$m
& $0.88$ & $1.25$ & $1.62$ & $2.20$ \\
Passband $\lambda_{FWHM}$, $\mu$m
& $0.22$ & $0.30$ & $0.20$ & $0.60$ \\
Spectral flux density from Vega, $10^3$Jy
& $2.55$ & $1.57$ & $1.02$ & $0.64$ \\
Spectral flux density from Vega, $10^9$photons/(m$^2\cdot$s$\cdot\mu$m)
& $43.73$ & $18.96$ & $9.50$ & $4.39$ \\
Sky brightness, $m$
& $20.00$ & $16.60$ & $14.40$ & $12.00$ \\
Sky brightness (continuum), $10^{-3}$Jy/arcsec$^2$
& $0.05$ & $1.50$ & $3.00$ & $15.00$ \\
Sky brightness (continuum),
$10^3$photons/(m$^2\cdot$s$\cdot$arcsec$^2\cdot\mu$m)
& $0.86$ & $18.11$ & $27.95$ & $102.90$ \\
\hline
\end{tabular}
\caption{IR fluxes of a $0^m$-magnitude A0 V-type star in the Earth's
atmospheric transparency windows; night-sky brightness according
to~\cite{1998NewAR..42..503B}}
\label{IK-filters}
\vspace{1em}
\end{widetbl}
In this mode (see Fig.~\ref{modes}), the diagonal mirrors, M,
are tilted at an angle of~$45\degr$ to the light beam, and a
cooled pupil stop, PS, is placed between them. For
the 6-m BTA telescope the scale in the plane of the
detector, D, is equal to $0.255''/$px and the field of view
has the size of $4\'3\times4\'3$. The main interference filters
(I, J, H, and K, and also filters for some molecular
bands), F, are located in the divergent beam in front
of the collimator. The image is focused by shifting the
detector, D.

The exit pupil is cooled down to a temperature
of about~150\,K in order to reduce thermal noise.
Filter wheels (located in the divergent beam) and slit
wheels (located in the focal plane of the instrument)
are kept at the temperature of the cryostat ``base''
with no special cooling applied. Diagonal mirrors are
coated with quartz-protected silver with a reflection
coefficient of no less than~98\% in the $0.8\div2.5\,\mu$m
wavelength interval.

To compute the limiting magnitude of the instrument in various parts of the
spectrum, one must know
the transmission of the ``instrument--telescope'' system, the spectral energy
distribution of the object
with atmospheric extinction taken into account, the
angular size of the turbulent disk of the star, and the
parameters of the detector.

We set the transmission of the instrument--telescope system equal to~$0.5$.

Table~\ref{IK-filters}  lists the extra-atmospheric energy distribution in the
spectrum of Vega, a $0^m$-magnitude~A0 V-type star, for different photometric
bands.

\begin{widetbl}
\begin{tabular}{|p{0.6\textwidth}|c|c|c|c|}
\hline
Band & I & J & H & K\\
\hline\strut
Pixel size, arcsec
&	 \multicolumn{4}{c|}{$0.254$}\\
Number of photons per second from Vega, $N_{\gamma0}$, $10^9$
& $125.08$ & $73.93$ & $24.71$ & $34.25$ \\
Number of electrons per pixel per second from Vega, $N_{e}$, $10^{9}$
& $5.15$ & $3.05$ & $1.02$ & $1.41$ \\
Number of electrons per pixel per second from the sky, $N_{sky}$, $10^3$
& $0.05$ & $0.70$ & $1.77$ & $22.36$ \\\strut
Detector noise during exposure  $T_{exp}=100\,$s, e$^-$
& \multicolumn{4}{c|}{$20.0$}\\
Estimated limiting magnitude (S/N$=5$), $m$
& $22.85$ & $20.90$ & $19.21$ & $18.19$ \\
Exposure for $N_\star-N_{b}=1000$, s
& $268.57$ & $74.99$ & $47.27$ & $13.35$ \\
Limiting magnitude ($T_{exp}=100\,$s and $N_\star-N_{b}=1000$), $m$
& $21.78$ & $21.21$ & $20.02$ & $20.37$ \\
\hline
\end{tabular}
\caption{Estimate of the limiting magnitude of the instrument when operating in
the photometric mode}
\label{photoE}
\end{widetbl}

The flux in the spectral band is $\Phi=\Phi_\lambda\Delta\lambda$,
where~$\Phi_\lambda$ and~$\Delta\lambda$ are the spectral flux
density and passband, respectively. We multiply the
flux by the effective area of the mirror $S=26\,$m$^2$ to
obtain the radiation power of the star in the given
spectral range. We then divide the power by the
average photon energy (which we assume to be
equal to the energy $h\nu$ of a photon of the central
wavelength of the filter passband) and multiply it by
the transmission $\eta_p$  of the optical path to obtain the
number of photons hitting the detector:
$$
N_\gamma = \frac{\Phi S}{h\nu}\eta_p.
$$

The image of an object of angular size $\alpha=1''$
in the primary focus of the 6-m BTA telescope has
the linear size of $L=F\cdot\alpha=116\,\mu$m, where~$F$ is the
focal distance of the telescope. If the optical parts of
the instrument do not affect the aperture ratio of the
beam arriving at the detector, such an object would
illuminate 31~pixels of the HAWAII-I array (with a
pixel area $S_{pix}=x^2=18.5\times18.5\,\mu$m$^2$). A factor of
K reduction of the focal ratio decreases the image
size by the same factor ($116/K$~$\mu$m/arcsec), and its
area in pixels by a factor of~$K^2$ . For~$K=1.6$ one
arcsecond in the sky corresponds to 3.9\,pixels on the
detector. We take the average quantum efficiency of
the HAWAII-I array equal to~$Q = 0.5$. The number of
electrons accumulated in one second by a single pixel
of the detector is evidently equal to:
$$
N_{e} = \frac{Q\cdot N_\gamma K^2 S_{pix}}{\pi (L/2)^2} =
\frac{4K^2QS_{pix}N_\gamma}{\pi F^2\alpha^2}.
$$

The sky background further reduces the signal-tonoise ratio. Let~$s$ be the sky
brightness (in photons/m$^2$/sr/s) and~$N_{CCD}$, the total detector noise
(in electrons/s). The star light collected by the telescope (proportional
to~$\Phi S$) produces $N_\star$~electrons
per second in the quantum well of a single pixel.
Sky background radiation produces $N_{sky}$~electrons
in the same way (the sky flux is $\Phi_{sky}=s\beta^2$, where
$\beta$~is the angular size of the given sky area). To these
we should add thermal electrons, $D$. As a result, we
obtain: $N = N_\star + N_{sky} + N_{CCD}$.

In the case of Poisson statistics the signal-to-noise ratio of the system is
$B=N_\star/\sqrt N$, from which
we determine the minimum flux from the star de-
tectable at a given~S/N:
$$
N_\star^2-B^2N-B^2(N_{sky}+N_{CCD}) = 0.
$$
It follows from this that
$$
N_\star =
\frac{B^2}{2}\left(1+\sqrt{1+\frac{4(N_{sky}+N_{CCD})}{B^2}}\right).
$$

\begin{widepict}
\includegraphics[width=0.8\textwidth]{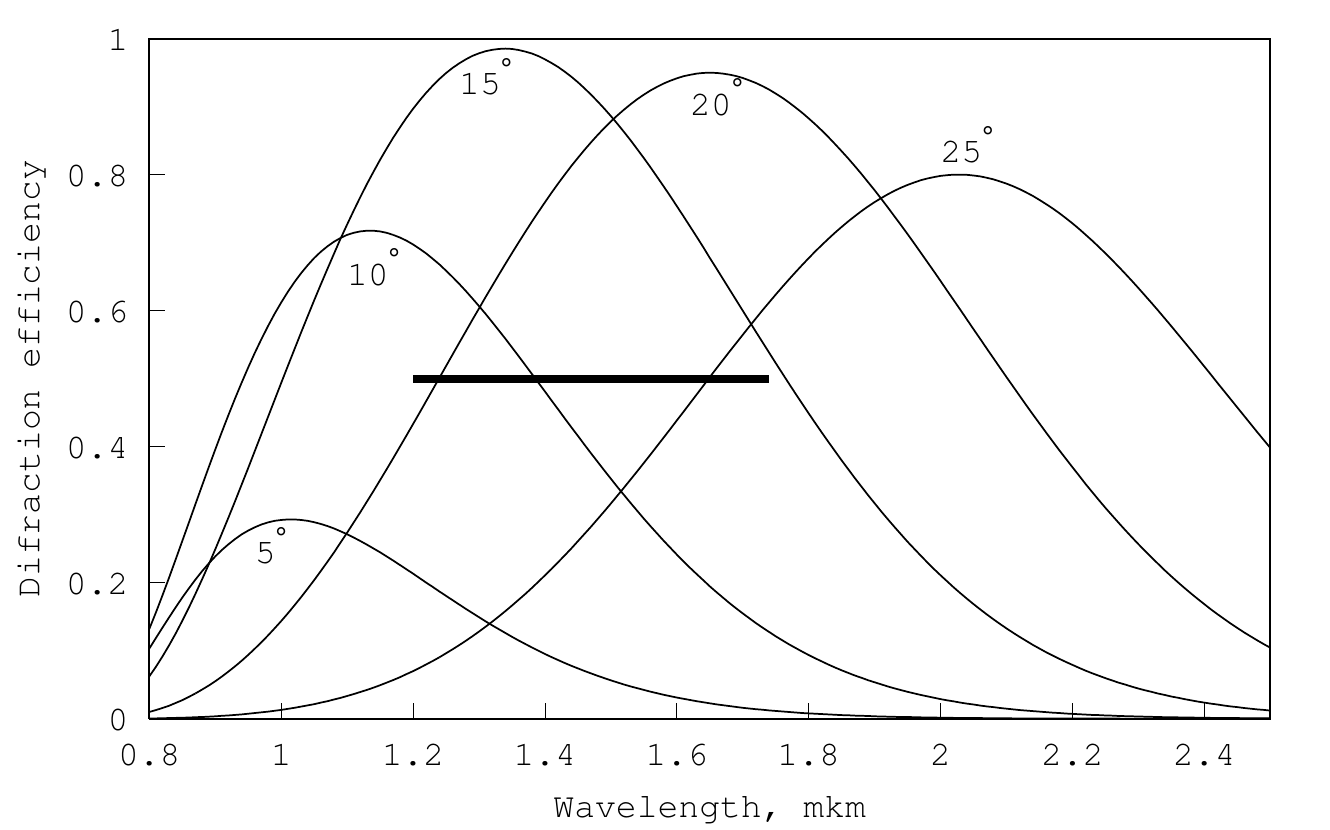}
\caption{Diffraction efficiency as a function of wavelength.
The bold line indicates the frame size.}
\label{eta_B}
\end{widepict}

We can estimate $N_{sky}$ in the given photometric band by the formula
$$
N_{sky} = N_0\cdot10^{-0.4m_{sky}},
$$
where $m_{sky}$ is the surface brightness of the sky (in
magnitudes per squared arcsecond) and~$N_0$ is the
number of photoelectrons per pixel from an extended
source with a surface brightness of~$0^m$/arcsec$^2$.

According to the manufacturer's data, the array
has a dark current of less than~$0.1\,$e$^-$/s and a readout
noise of less than~$10\,$e$^-$ (readout noise depends on the
number of intermediate nondestructive signal readouts during exposure). We then
estimate the limiting
magnitude of the instrument for photometric observations by the formula
 $\Delta m = -2.5\lg(N_\star/N_0)$ (see Table~\ref{photoE}).

\subsection{Spectroscopic Mode}
In this mode the cooled pupil stop is replaced by
a VPHG (see Fig.~\ref{modes}) in such a way that the turnable mirrors produce
incidence angles on the grating
that satisfy the Bragg condition for the given central
wavelength. In this case, a set of slits is placed in the
focal plane. Figure~\ref{eta_B} shows how the spectral interval
is selected by varying the Bragg angle by rotating the
diagonal mirrors.

We now use equations from~\cite{2000PASP..112..809B, 2004PASP..116..403B} to
estimate the parameters of the volume phase holographic grating (VPHG).

\begin{widetbl}
\begin{tabular}{|p{0.65\textwidth}|c|c|c|c|}
\hline
Band & I & J & H & K\\
\hline
Number of electrons per pixel per second $N_{e}$, $10^6$
& $6.15$ & $2.67$ & $1.34$ & $0.62$ \\
\hline
\multicolumn{5}{|c|}{Continuum}\\
\hline
Number of electrons per pixel per second from the sky, $N_{sky}$
& $0.05$ & $1.05$ & $1.62$ & $5.95$ \\
Detector noise during exposure  ($T_{exp}=1000\,$s), e$^-$
& \multicolumn{4}{c|}{110.0}\\
Estimated limiting magnitude, $m$ (S/N$=5$)
& $19.76$ & $17.91$ & $16.96$ & $15.47$ \\
Exposure for  $N_\star-N_{b}=1000$, $10^3$ss
& $13.01$ & $5.46$ & $4.53$ & $2.49$ \\
Limiting magnitude ($T_{exp}=1000\,$s and $N_\star-N_{b}=1000$), m
& $16.97$ & $16.06$ & $15.32$ & $14.48$ \\
\hline
\multicolumn{5}{|c|}{OH lines}\\
\hline
Number of electrons per pixel per second from the sky, $N_{sky}$
& $0.08$ & $1.48$ & $8.43$ & $66.48$ \\
Estimated limiting magnitude  (S/N$=5$), m
& $19.68$ & $17.75$ & $16.12$ & $14.19$ \\
Exposure for  $N_\star-N_{b}=1000$, $10^3$s
& $12.09$ & $4.71$ & $2.11$ & $0.77$ \\
\hline
\end{tabular}
\caption{Estimate of the limiting magnitude of the instrument when operating in
the spectroscopic mode}
\label{spectraE}
\end{widetbl}

The diffraction efficiency of the VPHG in the
planes parallel (the $P$-plane) and perpendicular (the $S$-plane) to the
direction of the periodic structure of the
grating can be estimated by the following formulas:
$$
\eta_s = \sin^2\Bigl(\frac{\pi\Delta n_g d}{\lambda\cos\alpha_g}\Bigr),\quad
\eta_p = \sin^2\Bigl(\frac{\pi\Delta n_g
d}{\lambda\cos\alpha_g}\cos(\alpha_g+\beta_g)\Bigr),
$$
where $d$~is the thickness of the grating; $\Delta n_g$~is the
modulation amplitude of the grating refractive index~$n_g$
(often $n_g(z)=n_g+\Delta n_g\cos[f(z)]$); $\alpha_g$~is the
light incidence angle onto the grating plane of uniform~$n_g$
(inside the grating the angle is counted from
the plane of discontinuity), and $\beta_g$~is the diffraction
angle (inside the grating). If $\alpha_g=\beta_g$ and the Bragg
condition
$$
m\nu_g\lambda = 2n_g\sin\alpha_g,
$$
where  $\nu_g$~is the modulation frequency of the grating
refractive index, is satisfied, we obtain for the simplest
case, when the lines are perpendicular to the grating
plane and $\alpha_g=\beta_g$:
$$
n_g\sin\alpha_g = n_{air}\sin\alpha,\quad
\eta_s = \sin^2\Biggl(\frac{\pi n_g\Delta n_g
d}{\lambda\sqrt{n_g^2-\sin^2\alpha}}\Biggr),
$$
$$
\eta_p = \sin^2\Biggl(\frac{\pi n_g\Delta n_g
d}{\lambda\sqrt{n_g^2-\sin^2\alpha}}\Bigl(1-\frac{2\sin^2\alpha}{n_g^2}
\Bigr)\Biggr).
$$

The wavelength of maximum diffraction efficiency for
the $\beta_g=\alpha_g$~case can be estimated by the following
formula:
$$
\lambda_{max}=\frac{2n_g\Delta n_g d}{(1+2n)\sqrt{n_g^2-\sin^2\alpha}},\qquad
n=\overline{1,\infty}.
$$

We now compute the dispersion curve of the grating
by the formula $\alpha=\arcsin(m\nu_g\lambda/2)$, where $m$~is
the diffraction order. We thus derive the following
formula for the reciprocal linear dispersion:
$$
\frac{d\lambda}{dx}=\frac{2\Lambda\cos\alpha}{F_c}\biggl[
1 - \Bigl(\frac{\lambda}{2\Lambda}\Bigr)^2\biggr],
$$
where  $F_c$~is the focal distance of the focal reducer
camera.

The angular width of the spectrum is
$\Delta\alpha=\alpha_{red}-\alpha_{blue}$, where
$\alpha_{blue}$ and~$\alpha_{red}$ are the blue
and  red boundaries of the spectrum, respectively. The
linear width of the spectrum is determined by the focal
distance of the camera:
$l=F_c(\tan\alpha_{red}-\tan\alpha_{blue})$.
To efficiently use the detector surface, we must
choose the focal distance of the camera based on the
optimum relationship between the spectral resolution
of the instrument (i.e., an element of the spectrum
should occupy at least two to three pixels) and the
halfwidth of the diffraction efficiency of the grating.
 
To make a spectrograph with a resolution of $R\sim3000$ at
$\lambda\sim1.65\,\mu$m, the average size of the
resolution element should be $\delta\lambda=1.65/3000=0.55\,$nm.
For the first order of diffraction the average angular
width of the element of resolution is $\delta\alpha=\nu_g\delta\lambda$.
For example, for a 600~lines/mm grating 
($\nu_g=6\cdot10^{5}$m$^{-1}$), this parameter is equal to
$\delta\alpha=3.3\cdot10^{-4}\text{rad}=68''$. ``Placing'' one element
of resolution in two detector pixels,
$N_W=2$ ($w=37\,\mu$m), we estimate the focal distance
of the camera: $F=11.2\,$mm. In the case of a factor of
$K = 1.6$ focal ratio reduction the angular width of the
slit is~$wK/F_{tel}=0\"51$.

To estimate the limiting magnitude of the
spectrograph, we compute the average flux in atmospheric
windows per spectrograph resolution element (by
multiplying the flux density by the width of the resolution element).

For the 6-m BTA telescope the height of the
spectrum of a star with a turbulent disk of angular
diameter~$1''$ is $N_H=116/18.5/1.6=3.9\,$px. A $0\"5$~wide
slit blocks~40\% of the star's image area, i.e., the flux
per spectral resolution element should be multiplied
by a factor of about~$0.6$ (if we ignore the Gaussian
distribution of brightness over the turbulent disk of
the star). The fraction of the area of a stellar image of
diameter~$D$ that passes through a slit of width~$h$ can
be determined by the formula
$$
\eta_{slit} = 1 - \frac2{\pi}\left(\arccos\frac{h}{D} -
\frac{h}{D}\sqrt{1-\frac{h^2}{D^2}}\right);
$$

Thus the power per one spectral resolution
element outside the instrument is spread over
$N_{pix}\sim8\,$px. Given the transmission of the optics
($\eta_P\approx0.5$), the average diffraction efficiency of the
grating ($\eta_G\approx0.5$), and the cutoff of a part of the
image of the turbulent disk of the star at the slit
($\eta_S\approx0.6$), we compute the average number of
photoelectrons per detector pixel per second as:
$$
N_{e} = \frac{\Phi_\lambda \delta\lambda \cdot\lambda}{hc} \cdot
\frac{SQ\eta_P\eta_G\eta_{slit}}{N_{pix}},
$$
where $\Phi_\lambda$ and~$\lambda$ are the spectral flux density of the
star's radiation and the mid-wavelength of the
spectral interval considered, respectively.

We can estimate in the same way the limiting
magnitude for spectroscopy in the case of a 1000\,s
long exposure. We estimate the number of photoelectrons
created as a result of the continuum sky
brightness by the formula
$$
N_{e,\,cont} = \frac{c}{\lambda^2}\frac{\Phi_{\lambda,\,sky}\lambda}{hc}
\frac{\delta\lambda S \eta_P\eta_G Q \rho}{N_W} =
\frac{\Phi_{\lambda,\,sky}}{h\lambda}\frac{\delta\lambda S \eta_P\eta_G Q
\rho}{N_W},
$$
where $\rho$~is the pixel area in sky steradians. The
corresponding estimates are listed in Table~\ref{spectraE}.

\begin{widepict}
\centerline{\includegraphics[width=0.6\textwidth]{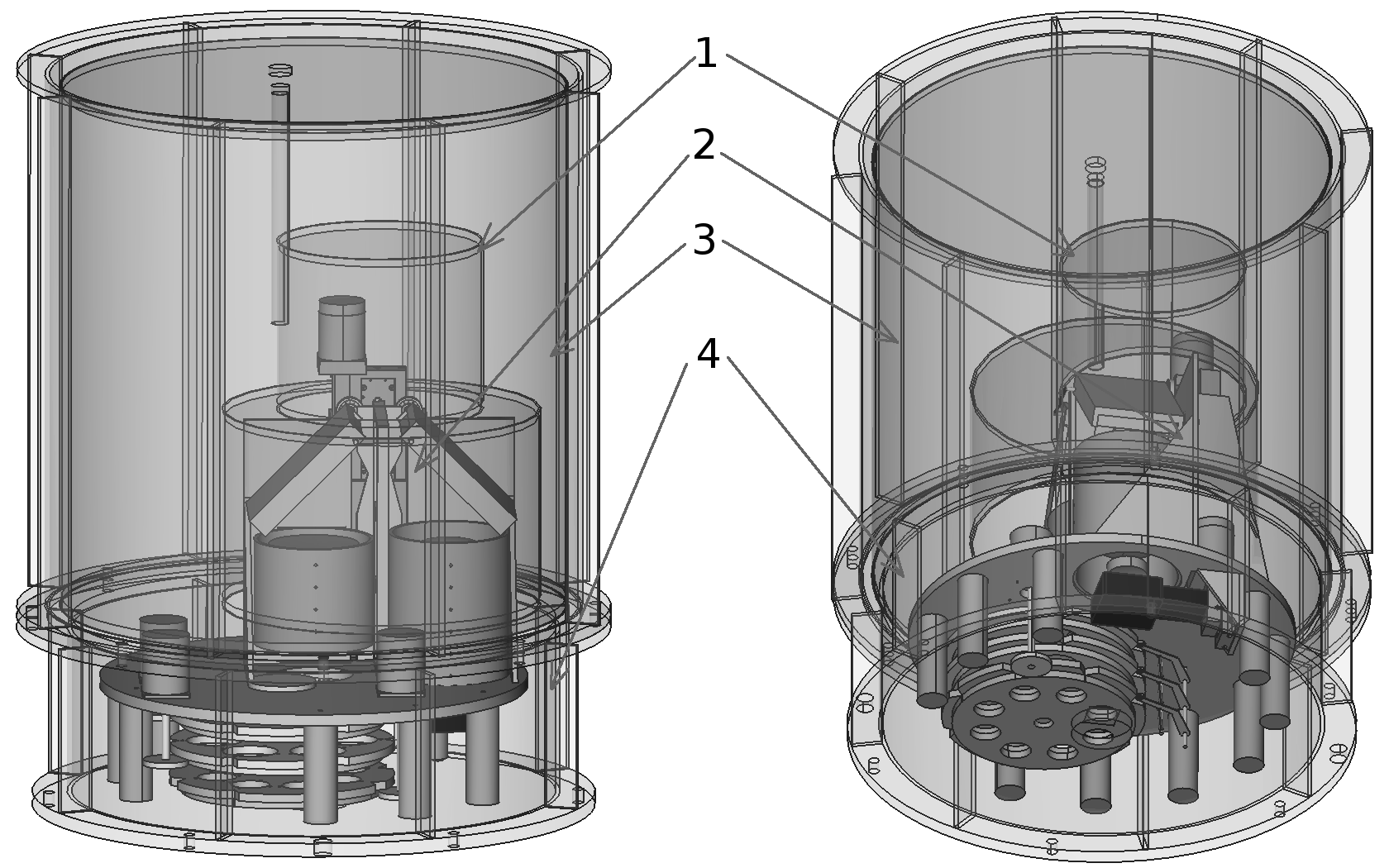}}
\caption{General view of the cryostat of the IR spectrometer: 1~--~liquid
nitrogen container; 2~--~main frame; 3~--~the top part of
the cryostat cabinet, and 4~--~the bottom part of the cryostat cabinet}
\label{cryo}
\end{widepict}

\section{Design of the Instrument}
\subsection{Cryostat}
The entire instrument is housed in a cylindrical
cryostat (see Fig.~\ref{cryo}) with the diameter and length
of no greater than~550 and~650\,mm, respectively. A
part of the cryostat is occupied by the no-spill
liquid-nitrogen container, made for the use in the primary
focus of the 6-m BTA telescope. The cryostat was
designed in accordance with the underlying principles
of the cryostats made earlier for the 6-m BTA
telescope~\cite{2001BTA_cryo}. Only two parts of the optical scheme
require deep cooling: the detector and the exit pupil unit
with diffraction gratings. Highly stable temperature is
needed only for the detector.

Let us refer to the cryostat part that houses the
flange with the optical window as the bottom part.
The vacuum chamber consists of two asymmetrically
arranged cylindrical containers: the lower container is
located in the bottom, detachable part of the cryostat,
and the upper container is located inside the
liquid-nitrogen container and has the form of a cylindrical
depression in the container.

The flanges of the top and bottom parts of the
cryostat facilitate its assembling and disassembling:
by opening the top cap with the liquid-nitrogen
container we get access to the entire internal
arrangement of the cryostat.

The case of the vacuum chamber is made of
titanium sheet. To reinforce the external walls of the
cryostat and enhance their bearing capacity, rib
stiffeners aligned along the vertical axis are welded to the
external walls of both parts of the cryostat.

The shell ring of the lower chamber houses
SNTs-type electric pressure seal connectors for the
input/output of control signals and supply of power to
the internal elements; windows facilitating assembly
and disassembly of the entire structure, and the
exhaust tube equipped with a safety valve.

A shutter is placed in front of the ``warm'' entrance
window of the cryostat at the front flange. All op-
tical and electromechanical equipment is located on
a stainless steel welded frame mounted on the lower
flange of the instrument via heat-insulating pads (see
Fig.~\ref{mount_parts}).

The cooled wheel with slits is located in the focal
plane of the telescope. Behind this wheel two filter
wheels are located in the divergent beam. These are
followed, along the optical axis, by the collimator that
forms the exit pupil of the instrument. Behind the
collimator, two diagonal mirrors that turn the light beam
by~$180\degr$ are located. A diffraction grating is placed
between the diagonal mirrors in the spectroscopic mode,
and a cooled pupil stop~--- in the photometric mode.
The objective followed by the detector are located
behind the diagonal mirrors.

\begin{widepict}
\centerline{\includegraphics[width=.7\textwidth]{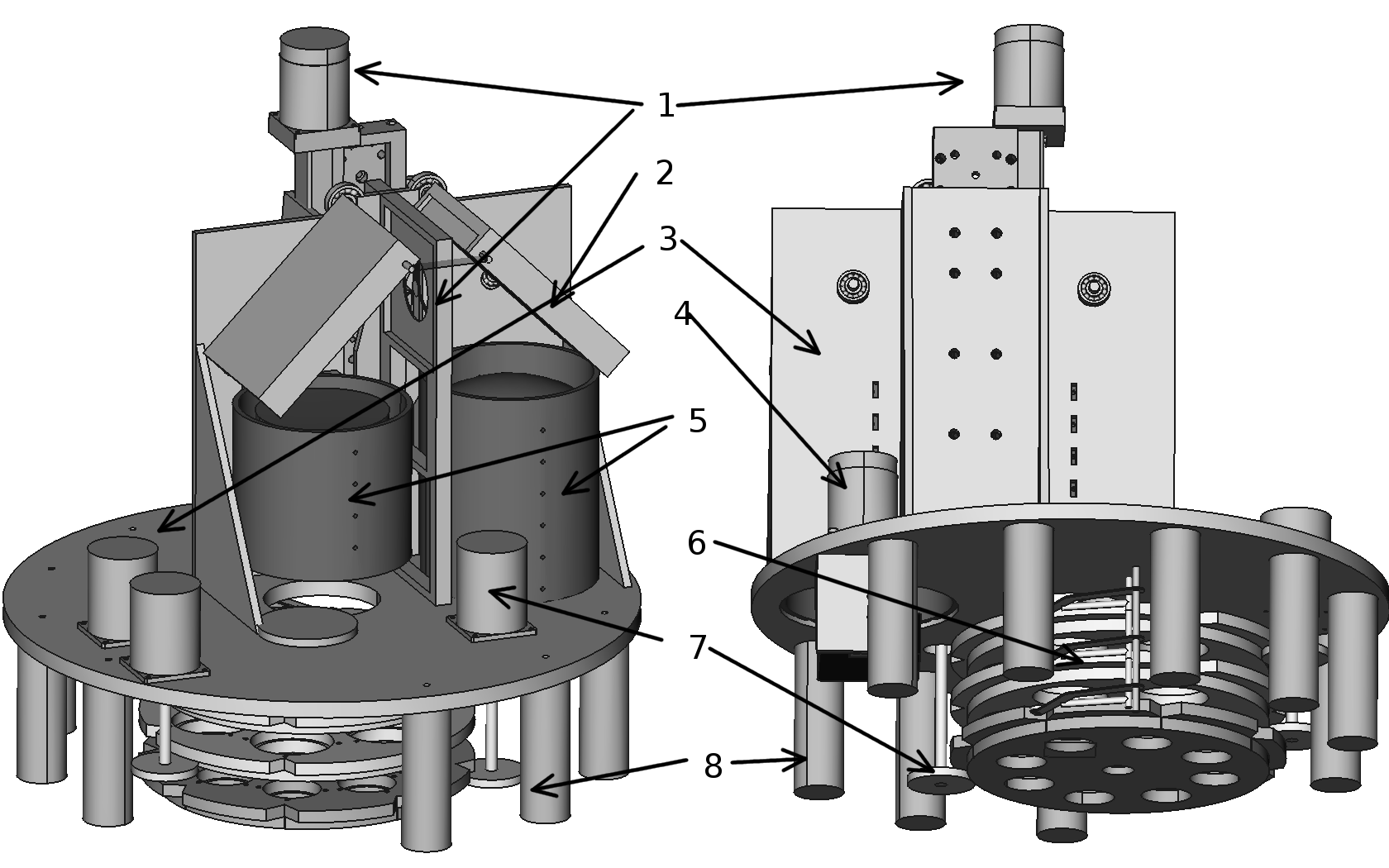}}
\caption{General view of the cryostated IR spectrometer: 1~--~linear stage with
the slit and pupil stop unit; 2~--~diagonal mirrors;
3~--~mounting frame of the instrument; 4~--~focus stage; 5~--~collimator and
camera; 6~--~wheel unit; 7~--~wheel drives, and 8~--~heat-insulating supports}
\label{mount_parts}
\end{widepict}

The liquid-nitrogen container is located in the top
part of the cryostat and has the form of a cylinder
with asymmetric depression. The shell ring and the
top caps of the liquid-nitrogen container are made of
stainless steel. The bottom ring of the container with
heat-removing contact areas is made of copper.

The liquid-nitrogen container is fixed inside the
cryostat case by welding it to the inner filler tube
(which, in turn, is welded to the external nitrogen
tube, which is welded to the external flange of the
cryostat). In addition, the bottom part of the shell
ring of the liquid-nitrogen container is stretched with
kevlar threads to ensure the required stiffness of the
structure.

The principal elements of the optical scheme are
cooled by flexible copper cooling pipes mounted to the
copper rim of the liquid-nitrogen container. To ensure
the operation of the cryostat at any tilt angle, the caps
of the container are connected inside it with copper
cooling pipes.

The slit wheel has eight slots. Seven of these slots
are occupied by slits and the eighth slot is left
unoccupied to facilitate the photometric mode of
operation. Both filter wheels have six slots with five slots
occupied by filters and the sixth left free. The wheels
are located on a common axis rigidly mounted on the
main plate of the supporting frame. Each wheel is
mounted on a pair of abutted bearings. The stepper
motors that rotate the wheels are fixed on the same
plate. Torque is transferred via gears mounted on
engine shafts; the other end of each gear shaft is left
loose.

To reduce the cost and mass of the drive of
diagonal mirrors and to simplify the synchronization
of mirror rotation with the translational motion of
the shifter with the grating unit and pupil stop, the
translational motion of the shifter is transformed into
rotational motion of mirrors via a sine drive
mechanism. The mirrors are held by shaft axles fixed by a
pair of abutted bearings.

The mechanics of the instrument are driven by
standard devices: the linear stage \hbox{8MT173V-10}
(Standa) is used for focussing (moving the detector);
the linear stage \hbox{8MT175V-150} (Standa) moves the
set of elements in the exit pupil and ensures the
operation of the sine-drive mechanism; the stepper
motors \hbox{VSS42} drive the wheels.

The estimated mass of the instrument when
assembled and nitrogen-filled is about~110\,kg.

\subsection{Sine-Drive Mechanism}
The sine-drive mechanism consists of gimbal-mounted
mirrors and a symmetric cam rigidly fixed
onto the linear translator of the pupil stop and
diffraction gratings. To reduce friction, the mirrors and the
cam are coupled via bearings.

We modelled the sine-drive mechanism to
determine the optimum positions for the mirror rotation
axis, the axis of the thrust bearings, and the shape
of the cam, and thereby minimize the offset of the
exit pupil relative to the central plane of the shifter
resulting from a change of the mirror tilt angle, and
minimize the offset of the beam center relative to the
center of the corresponding optical element on the
shifter.

Figures~\ref{sinmech-UO} and~\ref{sinmech} show
the notation used in the
model. Let us assume that the coordinates of the
mirror rotation axis (point~$O'$) relative to the mirror
center are $(x', y')$, and those of the axis of the thrust
bearing (point~$O''$) are $(x'', y'')$. We denote the radius
of the bearing by~$R$. We count the mirror tilt angle, $\alpha$,
from the horizontal direction and denote the cam apex
angle by~$\gamma$.

As a result of the vertical displacement of the cam
by~$dZ_s$ the mirror tilt angle changes and the center of
the reflected light beam shifts by~$dZ_r$. The coordinates
of point~$O''$ relative to point~$O'$ can be determined
from Fig.~\ref{sinmech-UO}:
\begin{equation}
\left\{
\begin{aligned}
x_{O''} &= (x'' - x')\cos\alpha - (y'' - y')\sin\alpha;\\
y_{O''} &= (x'' - x')\sin\alpha + (y'' - y')\cos\alpha.
\end{aligned}\right.
\label{Ocoords}
\end{equation}

Point $M$ where the bearing touches the cam lies on
the normal to the cam that passes through point~$O''$.
As the cam moves, this point moves over its surface
so that vector~$\mathbf{O''M}$ undergoes parallel displacement.
The coordinates of point~$M$ relative to point~$O'$ are:
\begin{equation}
\left\{
\begin{aligned}
x_M &= x_{O''} + R\cos\gamma;\\
y_M &= y_{O''} - R\sin\gamma.
\end{aligned}\right.
\label{Mcoords}
\end{equation}

By tilting the mirror from $0\degr$ to angle~$\alpha$ we increase by
$\delta l = -x'\tan\alpha$ the length of the central ray incident
on the mirror. In the case of~$\alpha=0\degr$, the $Y$~coordinate
relative to the mirror plane of the point of incidence of
the reflected ray on the diffraction grating is equal to
$y_\mathrm{ray} = \delta l - w\cot 2\alpha$.

The vertical offset of the cam, up to a constant
term uniquely determined by the configuration of the
system, is related to the coordinates of the point where
the thrust bearing touches the cam:
\begin{equation}
Y_0 = \mathbf{C} + y_m + (w - x' - x_M)\cot\gamma,
\label{Ycoords}
\end{equation}

Thus equations~\eqref{Ocoords}, \eqref{Mcoords} and~\eqref{Ycoords}
allow deriving the dependence of the cam shift on the incidence
angle on the mirror. Inversion of this dependence yields
the analytical relation between the cam shift and the
incidence angle on the grating.

\begin{pict}
\includegraphics[height=0.8\columnwidth,angle=-90]{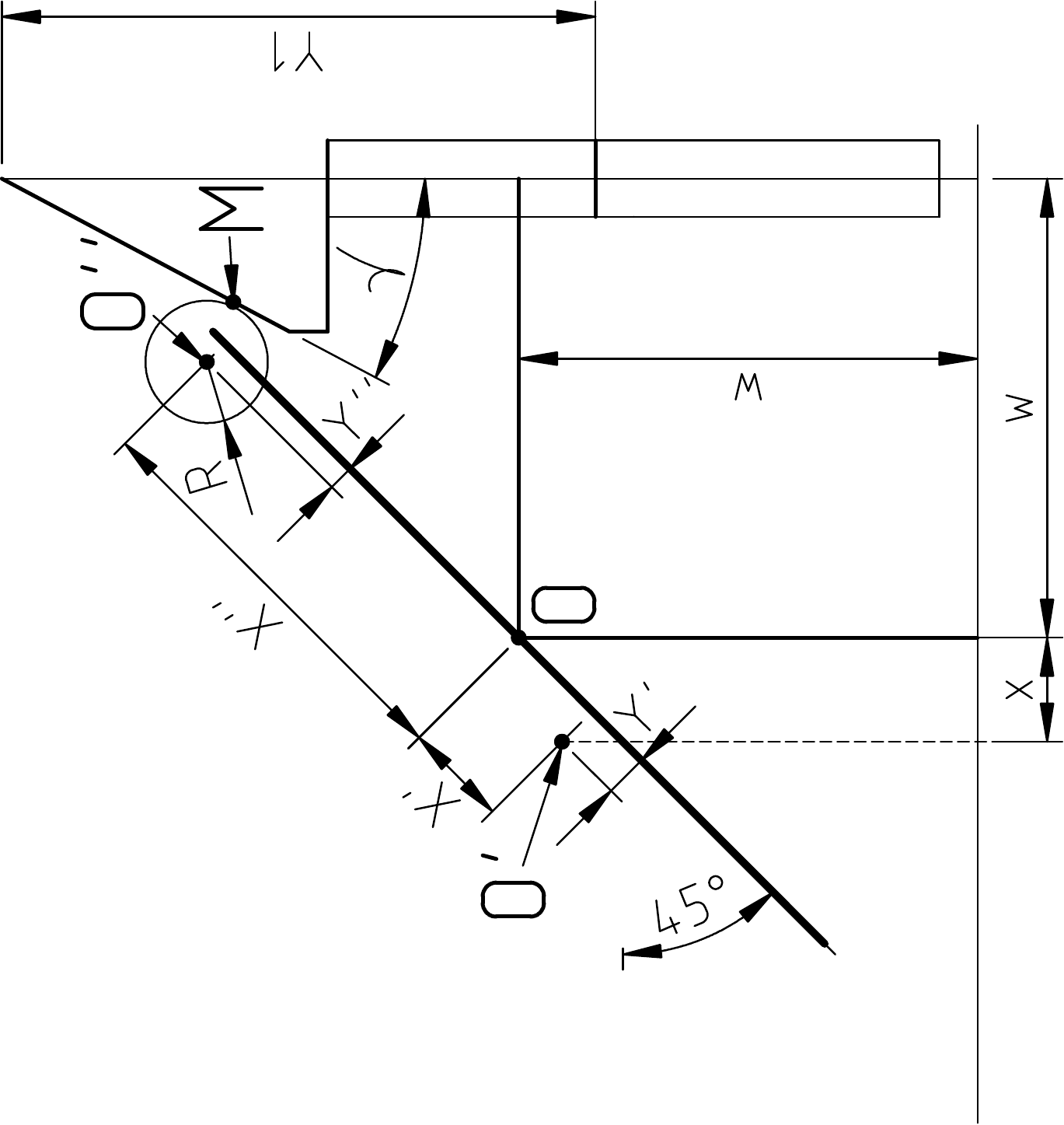}
\caption{Sine-drive mechanism notation}
\label{sinmech-UO}
\end{pict}
\begin{pict}
\includegraphics[height=0.8\columnwidth,angle=-90]{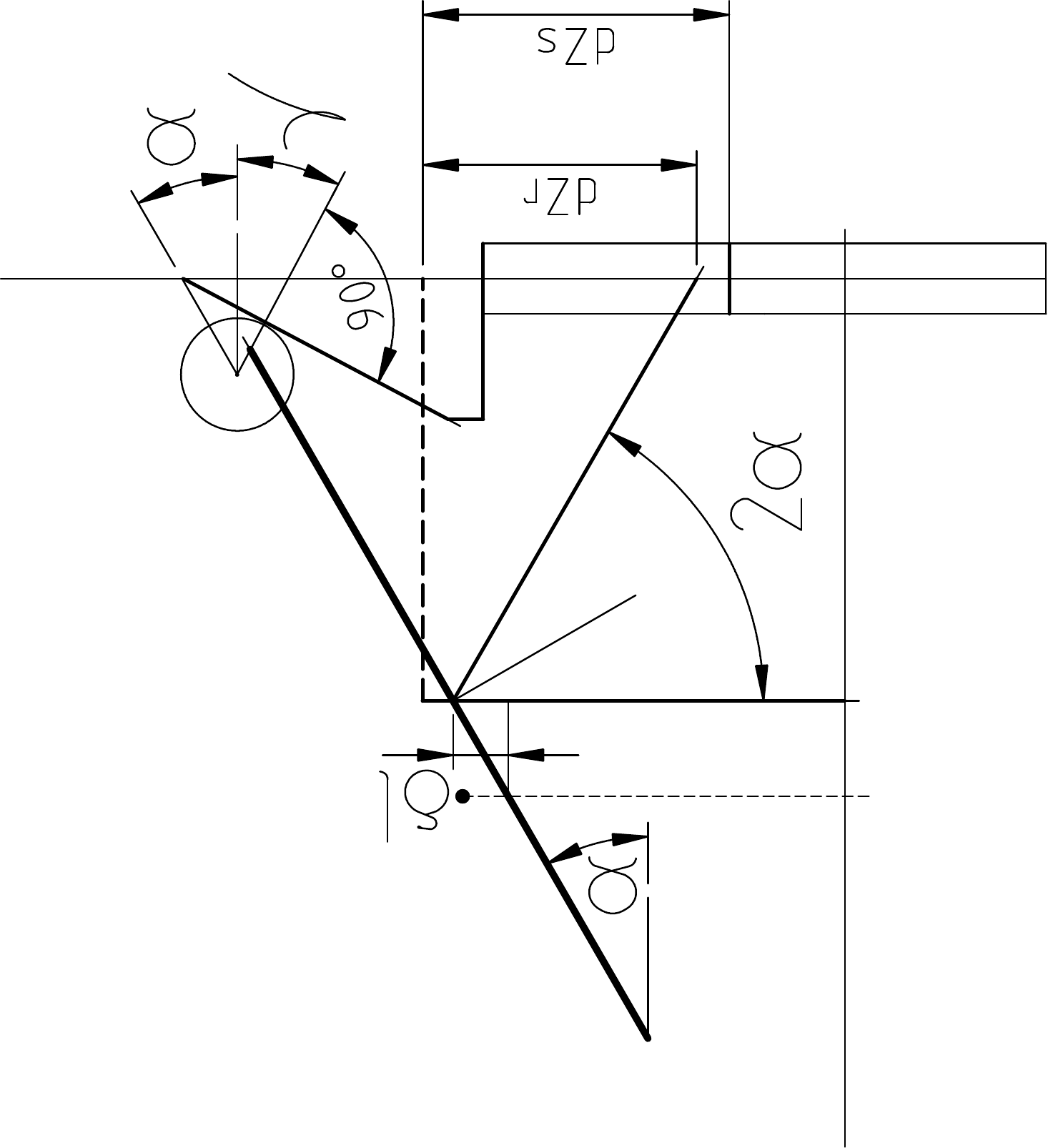}
\caption{Principle of operation of the sine-drive mechanism}
\label{sinmech}
\end{pict}

\subsection{Frame for Mounting the Elements of the Instrument}
All the optical and electromechanical equipment is
located on a monolithic stainless steel frame, which
is fixed to the bottom flange through heat insulating
(ceramic or woven-glass reinforced) joints. Figure~\ref{mount_parts}
shows the appearance of the frame with the elements
mounted onto it.

The frame is made by welding together its
structural elements, tempering the weld seams in a furnace
and subsequent polishing and boring on a
coordinate boring machine. The frame design ensures the
required stiffness while facilitating the assembly and
disassembly of its elements.

Instrument units are fixed to the frame with
screw joints. Electric wires are fixed by passing them
through the perforation in the frame walls and\slash or
with collars. Cooling pipes are fixed to the frame with
screw joints or braces through heat-insulating joints.

\subsection{Wheels}
The spectrometer scheme includes three wheels:
two filter wheels and one slit wheel.

The wheels are made of stainless steel and are
rotated by stepper motors. The current position of a
wheel is controlled using binary coding of the slots
with the Hall sensors.

The slit wheel has eight through holes with a
diameter of 35\,mm. The slit kit consists of round metal
frames with foil containing the slit of appropriate
width stretched over them. The frames are fixed to the
wheel surface with clamping plates. The slits occupy
seven slots of the wheel. The eighth slot is left free to
facilitate the photometric mode of operation.

Filter wheels have six slots with a diameter of
50\,mm. The filters are fixed with metal inserts and
clamping plates.

The wheels are located on a common axis that is
rigidly fixed to the main plate of the supporting frame.
The stepper motors that rotate the wheels are also
fixed onto the same frame. An additional plate fixes
the other end of the wheel axis. Each wheel is set on a
pair of abutted bearings. Torque is transferred via the
gears mounted on engine shafts; the other end of each
gear shaft is left loose.

\subsection{Optical Elements}
The collimator and camera are placed on the
vertical plate of the frame. The external mountings of
the collimator and camera have the form of
cylindrical structures with a 107-mm diameter. The crystal
lenses are fixed in the external mount with auxiliary
protective mounts that prevent damage to optics as a
result of tensions arising from temperature variations
(see Fig.~\ref{save_mount}).

By their design, the protective self-aligning mounts
balance the tensions arising as a result of thermal
deformations of mounts and lens material and stabilize
the position of the lens relative to the optical axis.
When the mount and lens temperature changes, the
difference of the lens and mount sizes is compensated
by moving the lens and the spring loaded retaining
ring along the optical axis of the instrument by sliding
the lens along the flats of the mount and retaining
ring. The motion of the retaining ring is made possible
by cylindrical guides. Protective mounts allow the
optics to operate with a temperature change rate of up
to $15\degr$C/hour (the recommended rate is $6\degr$C/hour).

\begin{pict}
\includegraphics[height=0.6\columnwidth,angle=-90]{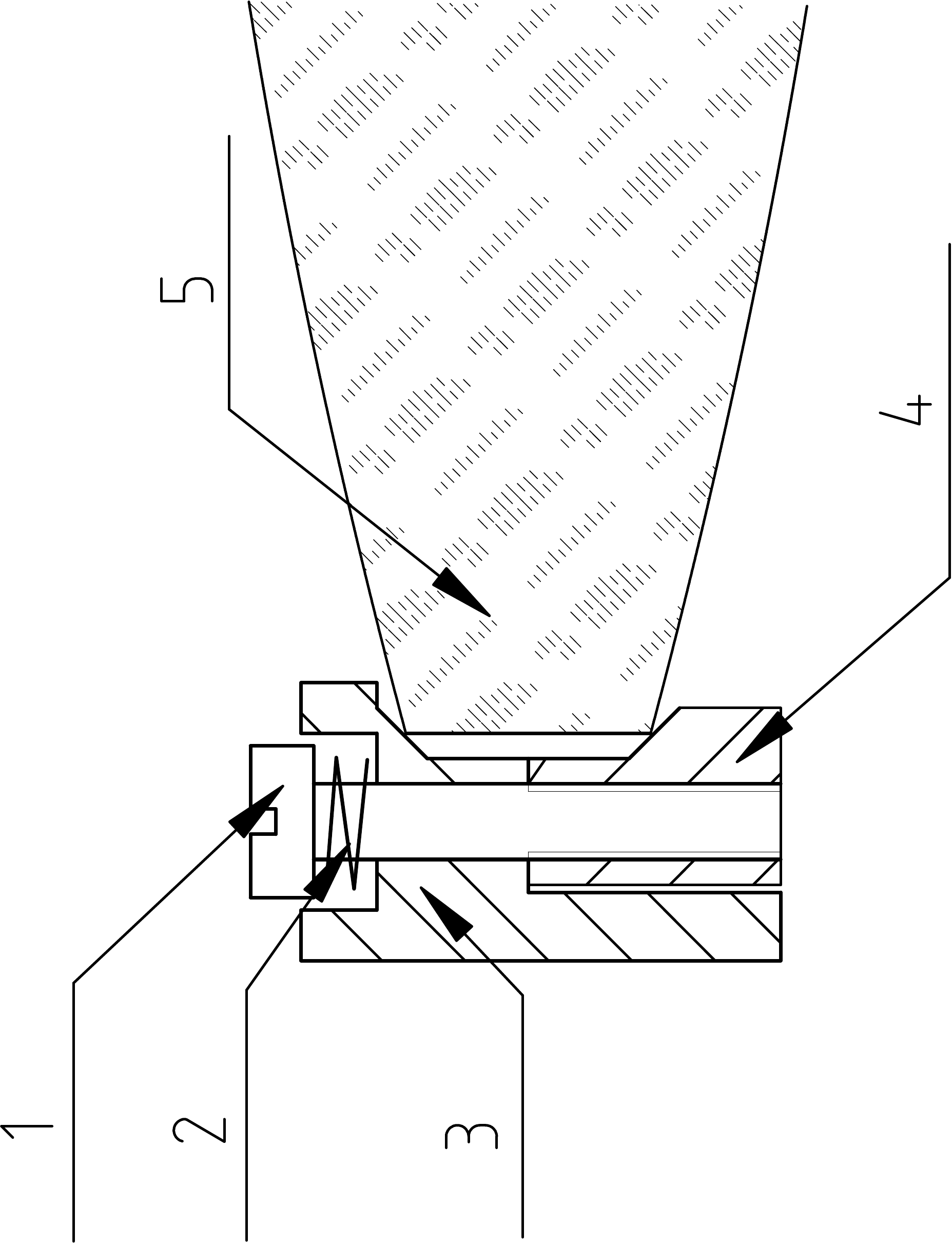}
\caption{Protective mounts of crystal lenses: 1~--~screw;
2~--~spring; 3~--~fixed part of the mount; 4~--~moving part of
the mount, 5~--~lens}
\label{save_mount}
\end{pict}

The external mounts of the optics are fixed to the
frame with screws. To ensure unique correct fixing
of the mounts, the frame has arched sockets.
Flexible cooling pipes are fed to the external mounts to
cool the optics down to the operating temperature
($-40\degr$C).

\subsection{Diagonal Mirrors}
The instrument is equipped with two tilting
mirrors to reduce the length of the optical system and
allow changing the incidence angle of the light beam
on the diffraction grating when operating in the
spectroscopic mode. When the instrument is operated in
the photometric mode, the mirrors are tilted by $45\degr$
relative to the light beam. When operating in the
spectroscopic mode the mirrors change the tilt angle
by~$30\degr$ to~45$\degr$  (in this case the sine-drive mecha-
nism is used to synchronize the displacement of the
diffraction grating and to transform the translational
motion of the grating and pupil stop unit translator
into mirror rotation).

The mirrors have the form of rectangular plates
made of sitall or another material whose reflective
surface is coated with quartz-protected silver. The
mirrors have overall dimensions of $120\times80\,$mm and
a thickness of 15\,mm. They are fixed in
stainless-steel mounts with clamping plates and spring stops.
Mirror mounts are placed on the vertical support of
the spectrometer frame on shaft axles that are fixed by
a couple of abutted bearings.

We modelled the sine-drive mechanism. The
optimum choice of the position of the mirror rotation axis,
the axis of the thrust bearing, the tilt angle, and the
size of the pusher allows minimizing the offset of the
exit pupil from the central plane of the stage resulting
from a change of the mirror tilt angle. According to
the model, the offset of the light beam from the center
of the diffraction grating does not exceed $1.33\,$mm
and the offset of the exit pupil center relative to the
VPHG plane does not exceed $2.62\,$mm throughout
the entire operating range of incidence angles. For the
$26\degr$~cam apex angle of the sine drive mechanism the
light incidence angle on the grating depends almost
linearly on the vertical displacement of the translator.

\subsection{Control and Recording Systems}
{\it The control system} allows moving the elements
of the instrument to change the mode of its
operation, and provides the data on the position of these
elements and their temperature. In the test mode
the control system should be able to operate
autonomously maintaining at least minimum
functionality. In the normal mode of operation the control
system is connected to the supervisory computer via
USB, RS-232, or RS-485 interfaces. The system is
powered either by a separate power-supply unit or by
the power supply unit of the data acquisition unit.

Cryostat units do not require simultaneous
operation of several drives and therefore the stepper motors
are controlled commutatively: only one stepper motor
can operate at a time. The engines are controlled via
commercial stepper motor drivers.

The position of the elements is controlled by Hall
sensors and end switches. The wheels reference
positions are binary coded by the corresponding
arrangement of magnets (the Hall sensors are placed on the
stationary part) --- this method reduces the number of
sensors and thereby also the number of links and
processed signals.

{\it The registration system} uses a cooled hybrid
HgCdTe CMOS HAWAII multiplexer to perform
numerical recording of the image produced by the
optical system of the telescope and the
spectroscopic\slash photometric equipment. The registration
system is controlled by the same PC to which the control
system is connected.

The registration system controls the operation
modes of the detector, maintains stable operating
temperature, controls exposure, readout, and saving
of the images in FITS format.

\section*{Acknowledgments}
The team participating in the development and
making of the IR spectrometer is rather big. We are
grateful to all the members of this team:
\begin{itemize}
\item Specialists of the Advanced Design Laboratory
of the Special Astrophysical Observatory (Nizhnii
Arkhyz) directed by S.\,V.~Markelov, who develop
the registration system of the instrument;
\item Specialists of Opto-Technological Service Co.Ltd.
(St. Petersburg) directed by B.\,N.~Ostrun, who
make the optics of the collimator and the camera;
\item  Specialists of the Institute of Applied Physics of
the Russian Academy of Sciences and the non-
profit partnership ``Experimental Design Bureau
of High-Tech Development'' (Nizhnii Novgorod)
directed by V.\,F.~Vdovin and Yu.\,Ya.~Brodskii, who
develop the preliminary design of the cryostat and
mechanics of the instrument.

\end{itemize}

\end{document}